\documentclass[a4paper,11pt]{article}
\usepackage{mathrsfs}
\usepackage{amsfonts}
\usepackage{amsmath}
\usepackage{amssymb}
\usepackage{indentfirst}
\usepackage{cite}
\newcommand{\pa}{\partial}

\newcommand{\al}{\langle}
\newcommand{\ar}{\rangle}

\newtheorem{corollary}{Corollary}[section]

\newtheorem{proposition}{Proposition}[section]
\title{Neumann Type Integrable Reduction to\\ the Negative-Order Coupled Harry--Dym Hierarchy}
\author {Jinbing Chen\thanks{cjb@seu.edu.cn;}\\ 
School of Mathematics, Southeast University,\\
Nanjing, Jiangsu 210096, P.R. China\\
}
\date{}
\begin{document}
\maketitle
\begin{abstract}\noindent
Based on the Lax compatibility, the negative-order coupled Harry--Dym (ncHD) hierarchy depending upon one parameter $\alpha$ is retrieved in the Lenard scheme, which includes the two-component Camassa--Holm (2CH) equation as a special member with $\alpha=-\frac14$. By using a symmetric constraint, it is found that only in the case of $\alpha>1$ the ncHD hierarchy can be reduced to a family of backward Neumann type systems by separating the temporal and spatial variables on the tangent bundle of a unit sphere. The resultant backward Neumann type systems are proved to be completely integrable in the Liouville sense via a Lax equation. Finally, for $\alpha>1$, the relation between the ncHD hierarchy and the backward Neumann type systems is established, where the involutive solutions of backward Neumann type systems yield the finite parametric solutions to the ncHD hierarchy.\\
\end{abstract}
\section{Introduction}
\label{sec:1}
It has been observed that integrable peakon equations can be placed in the negative-direction of soliton hierarchies \cite{1,2,3,4,6}. There has been growing interest in the study of negative-order integrable systems that emerge from spectral problems \cite{7,8,9,22,23,Blaszak}. Quite often, most negative-order flows can only be realized in the form of nonlocal integrable nonlinear evolution equations (INLEEs). The cases where the negative-order flows can be expressed in the form of local INLEEs with physical backgrounds will be of special interest, such as the Camassa--Holm (CH) equation \cite{1}, the Degasperis--Procesi (DP) equation \cite{2}, and some others with cubic nonlinearity \cite{3,4,6}. Note that the negative-order INLEEs are intractable due to the characteristic feature of nonlocality \cite{Blaszak}. Compared with positive-order INLEEs, the study of negative-order INLEEs has received less attention. 
However, it is the nonlocality that greatly enhances the variety of negative-order integrable systems, such as the emergence of peakons, cuspons, kinks, and so on \cite{1,6,9,10,11}. In this paper, we report that a nonlocal INLEE can be reduced to a pair of local finite-dimensional integrable systems (FDISs) by separating its temporal and spatial variables, and thus the derivation of explicit solutions of negative-order INLEEs can be simplified to solving some FDISs of solvable ordinary differential equations.

\par
The CH equation
\begin{equation}\label{1.1}
m_t+\omega \tilde{u}_x+2m \tilde{u}_x+m_x\tilde{u}=0,\qquad m=\tilde{u}-\tilde{u}_{xx},
\end{equation}
arises as a model describing the unidirectional propagation of shallow water waves over a flat bottom \cite{1}, and that of axially symmetric waves in a hyperelastic rod \cite{13,14}. It is noteworthy that the CH equation not only possesses stable solitons if $\omega>0$ \cite{15} and peaked solitary waves (peakons) if $\omega=0$ \cite{1}, but also allows the presence of wave breaking \cite{12}. Soon later, the study of peakons yielded various generalizations and modifications of the CH equation \cite{3,4,6}. One of the multicomponent generalizations is the 2CH equation encountered in the framework of deformation of a hydrodynamic-type bi-Hamiltonian structure \cite{3},
\begin{equation}\label{1.2}
m_t+2m \tilde{u}_x+m_x\tilde{u}+\rho\rho_x=0,\quad \rho_t+(\rho \tilde{u})_x=0,\quad m=\tilde{u}-\tilde{u}_{xx}.
\end{equation}
This extension has been demonstrated to be relevant to the context of shallow-water theory \cite{16}. The 2CH equation has attracted much interest in the past few years 
\cite{9,16,17,18,19,20,21,22,23}. Regarding connections to other integrable systems, it has been found that the 2CH equation is related via a reciprocal transformation to negative-order Ablowitz--Kaup--Newell--Segur (AKNS) flows \cite{x1,x2}, from which some multikink solutions and traveling wave solutions have been abtained for the 2CH equation \cite{9,22,23,x5}. Different from the above treatments, the aim of the present work is to demonstrate that the 2CH equation can be embedded into the negative-order coupled Harry--Dym (ncHD) hierarchy.

\par
Based on isospectral natures \cite{24}, INLEEs always appear together with a hierarchy of high-order members in the sense of Lax compatibility, and they can be formally depicted by the Lenard operator pair $K$, $J$ and the Lenard gradients. Therein, the kernel of $J$ usually gives rise to a positive-direction hierarchy of INLEEs, while the kernel of $K$ results in a sequence of negative-order INLEEs \cite{x3}. Moreover, the positive- and negative-order integrable systems share the spatial part of Lax representations, the commutable phase flows, and a pair of bi-Hamiltonian operators \cite{x4}. On one hand, the exact solutions to INLEEs known so far all have a finite number of parameters, which means that they may satisfy FDISs of solvable ordinary differential equations upon canceling the existing parameters. On the other hand, it has been noticed that INLEEs can be represented as the compatibility condition of two spectral problems. Then, the finite-dimensional integrable reduction is motivated in view of the nonlinearization of Lax pairs \cite{25}. Forward FDISs have been used to deduce soliton solutions, quasi-periodic solutions, rogue periodic waves, and polar expansion solutions for the positive-order INLEEs in various cases \cite{27,28,29,30,32,33-1}. Referring to the fruitful application of forward FDISs, we turn our attention to the finite-dimensional integrable reduction for negative-order INLEEs, which will not only enrich the content of FDISs themselves \cite{x6,x7} but also provide an effective way to solve negative-order INLEEs.

\par
The primary purpose of this study is to reduce the ncHD hierarchy to a family of backward Neumann type systems and then present its finite parametric solutions. Let us briefly explain the treatment of the ncHD hierarchy \cite{Blaszak}. In accordance with the Lax compatibility \cite{24}, we begin with the cHD spectral problem \cite{34} to reconstruct the ncHD hierarchy in the Lenard scheme, and we further specify Lax representations of the ncHD hierarchy in terms of a set of backward Lenard gradients. From the nonlinearizaiton of Lax pairs \cite{25}, we point out that the ncHD hierarchy with a single parameter $\alpha>1$ is decomposed into a family of backward Neumann type systems by separating the temporal and spatial variables on the tangent bundle of a unit sphere, while finite-dimensional integrable reduction can not be implemented for the case of $\alpha\leq1$, at least for the real space. The reduced backward Neumann type systems are rewritten as canonical Hamiltonian equations under the Dirac-Poisson bracket. Using a special solution of the Lenard eigenvalue equation, we obtain a Lax matrix that provides a sufficient number of conserved integrals for the backward Neumann type systems. The Lax matrix is shown to be satisfied by a Lax equation along with the backward Neumann type flows, which implies that two arbitrary backward Neumann type flows are consistent on the tangent bundle of a unit sphere. The relation between the ncHD hierarchy and the backward Neumann type systems is specified for $\alpha>1$, where the involutive solutions of backward Neumann type systems lead to the finite parametric solutions for the ncHD hierarchy. Based on our technical treatment, it is seen that the problem of obtaining solutions for nonlocal INLEEs is transferred to that of solving some FDISs of the Neumann type, which enhances the efficiency of solving negative-order integrable systems.

\par
The present article is organized as follows. In Sec. 2, starting from the kernel of the Lenard operator $K$, we deduce the ncHD hierarchy together with the 2CH equation, and then specify its Lax integrability. In Sec. 3, an explicit formalism of decomposition is presented for the ncHD hierarchy. Section 4 bridges the gap between the ncHD hierarchy and the backward Neumann type systems that results in finite parametric solutions to the ncHD hierarchy for $\alpha>1$.

\setcounter{equation}{0}
\section{ncHD Hierarchy and Its Lax Integrability}
\label{sec:2}
In this section, using the backward Lenard gradients, we retrieve the ncHD hierarchy depending on one parameter $\alpha$ \cite{Blaszak} and then specify its Lax integrability. It will be seen that the first nontrivial member in the list takes the local shape, and can be transformed to the previously known 2CH equation for $\alpha=-\frac14$ \cite{3,9,16}. Our point of departure is the cHD spectral problem attached with a single parameter $\alpha\in\mathbb{R}$ \cite{34}, expressed as
\begin{equation}
\varphi_x=U\varphi, \quad
U=\sigma_2-(\alpha+\lambda u+\lambda^2v)\sigma_3,\quad
\varphi=(\varphi_1,\varphi_2)^T,
\label{2.1}
\end{equation}
where $\lambda$ is the spectral parameter, $u$ and $v$ are the potentials, and
$$
\sigma_1=\left(\begin{array}{cc}
1&0\\
0&-1\\
\end{array}\right),\quad
\sigma_2=\left(\begin{array}{cc}
0&1\\
0&0\\
\end{array}\right),\quad
\sigma_3=\left(\begin{array}{cc}
0&0\\
1&0\\
\end{array}\right).
$$
Consider the stationary zero-curvature equation with respect to the spectral problem in Eq. (\ref{2.1}),
\begin{equation}
V_x=[U,V],\quad  V=\left\{\begin{array}{l}
a\sigma_1+
b\sigma_2+c\sigma_3=\sum\limits_{j\geq 0}(a_j\sigma_1+
b_j\sigma_2+c_j\sigma_3)\lambda^{-j},\quad j\geq0,\\
-a\sigma_1-
b\sigma_2-c\sigma_3=-\sum\limits_{j\leq -1}(a_j\sigma_1+
b_j\sigma_2+c_j\sigma_3)\lambda^{-j},\quad j\leq-1,\\
\end{array}
 \right. \label{2.2}
\end{equation}
which can be reformulated as
\begin{equation}\label{2.3}\begin{array}{c}
a_{jx}=\alpha b_j+ub_{j+1}+vb_{j+2}+c_j, \quad b_{jx}=-2a_j, \quad
c_{jx}=-2\alpha a_j-2ua_{j+1}-2va_{j+2},\\
-(2\alpha\partial+\frac12\partial^3)b_j=(\partial u+u\partial)b_{j+1}+(\partial v+v\partial)b_{j+2},\quad \partial=\partial/\partial x, \quad j\in \mathbb{Z}.\\
\end{array}
\end{equation}
Let $a_0=b_0=0$, $c_0=v^{\frac12}$ be the initial datum, and $w=b_{-2}$ [which is related to the parameter $\alpha$ through the recursive formula (\ref{2.3})]. Using the recurrence chain (\ref{2.3}), $a_j$, $b_j$, and $c_j$ can be uniquely determined up to the constants of integration, for instance,
\begin{equation}\label{2.4}
\begin{array}{c}
a_1=0,\quad a_2=-\frac14v^{-\frac32}v_x,\quad a_3=\frac38uv_xv^{-\frac52}-\frac14u_xv^{-\frac32}, \quad a_{-1}=0,\quad a_{-2}=-\frac12 w_x,\\
b_1=0,\quad b_2=-v^{-\frac12},\quad b_3=\frac12uv^{-\frac32}, \quad b_{-1}=-1,\quad  b_{-2}=w,\\
c_1=\frac12uv^{-\frac12}, \quad c_{-1}=\alpha,\quad c_{-2}=u-\alpha w-\frac12w_{xx}.\\
\end{array}
\end{equation}

\par
To deduce the ncHD hierarchy, we now convert the recursive formula (\ref{2.3}) into a pattern of Lenard gradients $\{g_j\}\ (j\in\mathbb{Z})$ and the Lenard operator pair $K$ and $J$ as
\begin{equation}\label{2.5}
Kg_{j}=Jg_{j+1}, \quad g_j=(-b_{j+2},-b_{j+3})^T,\quad j\in\mathbb{Z},
\end{equation}
where
\begin{equation}\label{2.6}
K=\left(\begin{array}{cc}
0&-(2\alpha\partial+\frac12\partial^3)\\
-(2\alpha\partial+\frac12\partial^3)&-(\partial u+u\partial)\\
\end{array}\right),\quad J=\left(\begin{array}{cc} -(2\alpha\partial+\frac12\partial^3)&0\\
0&\partial v+v\partial\\ \end{array}\right)
\end{equation} are two skew-symmetric operators. It is known from Eqs. (\ref{2.4})-(\ref{2.6}) that
\begin{equation}\label{2.7}\begin{array}{c}
g_{-4}=\left(\begin{array}{c}
-w\\
1\\
\end{array}\right),\quad g_{-3}=\left(\begin{array}{c}
1\\
0\\
\end{array}\right),\quad g_{-2}=\left(\begin{array}{c}
0\\
0\\
\end{array}\right),\\
g_{-1}=\left(\begin{array}{c}
0\\
v^{-\frac12}\\
\end{array}\right),\qquad
g_0=\left(\begin{array}{c}
v^{-\frac12}\\
-\frac12uv^{-\frac32}\\
\end{array}\right),
\end{array}
\end{equation}
\begin{equation}\label{2.8}
\ker
J=\{\varrho_1g_{-1}+\varrho_2g_{-3}|\forall
\varrho_1,\varrho_2\in \mathbb{R}\},
\end{equation}
\begin{equation}\label{2.9}
\ker
K=\{\varrho_{-1}g_{-3}+\varrho_{-2}g_{-4}|\forall
\varrho_{-1},\varrho_{-2}\in \mathbb{R}\}.
\end{equation}
Clearly, up to the kernel of $J$, the recursive formula $g_{j+1}=J^{-1}Kg_{j}\ (j\geq-1)$ leads to the forward Lenard gradients $g_0,g_1,\cdots,$ etc., and up to the kernel of $K$, the recursive formula $g_{j}=K^{-1}Jg_{j+1}\ (j\leq-5)$ results in the backward Lenard gradients $g_{-5}, g_{-6}, \cdots,$ etc., which coincides with the recursive relation (\ref{2.3}).

Let $g_{-}$ be a generating function of the backward Lenard gradients $\{g_{-j}\}\ (j\geq3)$,
\begin{equation}
\label{2.10}
g_{-}=(g_{-}^{(1)},g_{-}^{(2)})^T=\sum_{j=1}^{n-1}g_{-j-2}\lambda^{j-n},\qquad n\geq3.
 \end{equation}
We postulate that the time-dependent $\varphi$ also solves the following spectral problem determined by the backward Lenard gradients:
\begin{equation}
\varphi_{t_{-n}}=V^{(-n)}\varphi,\quad V^{(-n)}=-\frac12\partial g_{-}^{(1)}\sigma_1+g_{-}^{(1)}\sigma_2
-(\frac12\partial^2 g_{-}^{(1)}+(\alpha+\lambda u+\lambda^2v)g_{-}^{(1)})\sigma_3, \quad n\geq3,\label{2.11}
\end{equation}
After a direct calculation, we arrive at the identity
\begin{equation}
U_{t_{-n}}=V^{(-n)}_x-[U,V^{(-n)}]=U_*[(K-\lambda J)g_{-}], \label{2.12}
\end{equation}
where
$$U_*[\xi]=\left.\frac{d}{d\varepsilon}\right|_{\varepsilon=0}U(u+\varepsilon
\xi_1,v+\varepsilon \xi_2),\quad \xi=(\xi_1,\xi_2)^T.$$
The combination of Eqs. (\ref{2.5}), (\ref{2.10}), and (\ref{2.12}) gives a sequence of negative-order integrable systems, i.e., the ncHD hierarchy in the two-component version \cite{Blaszak} formally expressed by 
\begin{equation}\label{2.13}
(u_{t_{-n}},v_{t_{-n}})^T=Jg_{-n-2},\qquad
n\geq3.
\end{equation}
It follows from Eqs. (\ref{2.3})-(\ref{2.6}) that the first INLEE in Eq. (\ref{2.13}) can be represented as the following local form ({\it up to a constant of integration}) by virtue of an additional potential $w$:
\begin{equation}\label{2.14}
u_{t_{-3}}=-2uw_x-u_xw+v_x,\quad v_{t_{-3}}=-2vw_x-v_xw,\quad u=(2\alpha+\frac12\partial^2)w,
\end{equation}
which are the compatibility conditions of the Lax pair (\ref{2.1}), and
\begin{equation}
\varphi_{t_{-3}}=V^{(-3)}\varphi,
\label{2.15}
\end{equation}
\begin{equation}\label{2.16}
V^{(-3)}=\frac12w_x\sigma_1+(\lambda^{-1}-w)\sigma_2+
\left(\frac12w_{xx}+\alpha\left(w-\lambda^{-1}\right)-u+\lambda(uw-v)+\lambda^2vw\right)\sigma_3.
\end{equation}
To fix $\alpha=-\frac14$ and also identify the constant of integration as zero, from Eqs. (\ref{2.3}) and (\ref{2.4}) we know that $u=-(1-\partial^2)w/2$.
Let $w\rightarrow 2\tilde{u}$, $v\rightarrow \rho^2$, $t_{-3}\rightarrow \frac12 t$. The first ncHD equation (\ref{2.14}) can be transformed to the following standard 2CH equation [or the equivalent form Eq. (\ref{1.2})] \cite{3,9,16}:
\begin{equation}\label{2.17}\begin{array}{c}
\tilde{u}_t-\tilde{u}_{xxt}+3\tilde{u}\tilde{u}_x-2\tilde{u}_x\tilde{u}_{xx}-\tilde{u}\tilde{u}_{xxx}+\rho\rho_x=0,\\
\rho_t+\rho\tilde{u}_x+\rho_x\tilde{u}=0.\\
\end{array}
\end{equation}

On the other hand, by using the trace formula \cite{35}, we compute
\begin{equation}\label{2.18}\left(\frac{\delta}{\delta u},
\frac{\delta}{\delta v}\right)^T{\rm tr}\left(V\frac{\partial U}{\partial\lambda}\right)=\left(\lambda^{-s}\frac{\partial}{\partial
\lambda}\lambda^s\right)\left({\rm tr}\left(V\frac{\partial U}{\partial u
}\right),{\rm tr}\left(V\frac{\partial U}{\partial
v}\right)\right)^T,
\end{equation}
where $\left({\delta}/{\delta u},{\delta}/{\delta v}\right)^T$ is the variational derivative and $s$ is an absolute constant to be determined.
Replacing $b$ with the expansion of $\sum_{j=1}^{+\infty} b_{-j}\lambda^j$ in Eq. (\ref{2.18}), we compare the coefficients of $\lambda$ on both sides of Eq. (\ref{2.18}) and obtain
\begin{equation}\label{2.19}
s=-1,\quad\left(\frac{\delta}{\delta u},\frac{\delta}{\delta v}\right)^T(ub_{-j-1}+2vb_{-j})=(j+1)(b_{-j-1},b_{-j})^T,\qquad j\geq1.
\end{equation}
It follows from Eq. (\ref{2.5}) that the ncHD hierarchy can be rewritten as the bi-Hamiltonian formula
\begin{equation}\label{2.20}
(u_{t_{-n}},v_{t_{-n}})^T=J\left(\frac{\delta}{\delta u},\frac{\delta}{\delta v}\right)^T\mathscr{H}_{-n}=K\left(\frac{\delta}{\delta u},\frac{\delta}{\delta v}\right)^T\mathscr{H}_{-n-1},
\qquad n\geq3,
\end{equation}
where
\begin{equation}\label{2.21}
\mathscr{H}_{-n}=-\frac{1}{n}(b_{-n}+2vb_{-n+1}),\qquad n\geq3.
\end{equation}

\setcounter{equation}{0}
\section{Nonlinearization of Lax Pairs}
\label{sec:3}
A natural way to solve INLEEs is to reduce their dimensions to yield simpler lower-dimensional equations and then try to solve the reduced lower-dimensional equations. The nonlinearization of Lax pairs \cite{25} provides an effective way to keep the integrability of INLEEs while reducing their dimensions. The subsequent objective is to generalize the nonlinearization of Lax pairs to negative-order INLEEs, which produces backward Neumann type systems, for the purpose of solving negative-order INLEEs.

\par
In doing so, we first introduce some notions and symbols for later use. Let $\lambda_1,\lambda_2,\cdots,\lambda_N$ be $N$ distinct nonzero eigenvalues and $(p_j,q_j)$ be the eigenfunction corresponding to $\lambda_j$ $(1\leq j\leq N)$. For brevity, we denote $\Lambda={\rm diag}(\lambda_1,\cdots,\lambda_N)$, $p=(p_1,p_2,\cdots,p_N)^T$, and $q=(q_1,q_2,\cdots,q_N)^T$. The standard symplectic structure in $\mathbb{R}^{2N}$ is assumed to be $\omega^2=dp\wedge dq$, and the Poisson
bracket of two smooth functions $f=f(p,q)$ and $g=g(p,q)$ is defined by \cite{36}
$$\{f,g\}=\sum\limits_{j=1}^N\left(\frac{\pa f}{\pa q_j}\frac{\pa g}{\pa
p_j}-\frac{\pa f}{\pa p_j}\frac{\pa g}{\pa q_j}\right)=\left\al\frac{\pa f}{\pa
q},\frac{\pa g}{\pa p}\right\ar-\left\al\frac{\pa f}{\pa p},\frac{\pa g}{\pa
q}\right\ar,$$
where the diamond bracket $\langle\cdot,\cdot\rangle$ represents the inner product
in $\mathbb{R}^{N}$.

\par
We put the spectral problem (\ref{2.1}) into the component form
\begin{equation}\left(\begin{array}{c}
p_j\\
q_j\\
\end{array}\right)_x=\left(\begin{array}{cc}
0&1\\
-\alpha-\lambda_ju-\lambda_j^2v&0\\
\end{array}\right)\left(\begin{array}{c}
p_j\\
q_j\\
\end{array}\right) ,\qquad 1\leq j\leq N.\label{3.1}
\end{equation}
It is known from \cite{25} that the functional gradient of $\lambda_j$ with respect to the potentials $u$ and $v$ is
\begin{equation}
\nabla \lambda_j=(\delta\lambda_j/\delta u,
\delta\lambda_j/\delta v)^T=(\lambda_jp_j^2,\lambda_j^2p_j^2)^T,
\label{3.2}
\end{equation}
which solves the Lenard eigenvalue equation
\begin{equation}
(K-\lambda_j J)\nabla \lambda_j=0. \label{3.3}
\end{equation}
We place the following constraint on the relationship between $g_0$ and the sum of $\nabla \lambda_j$ from $1$ to $N$ \cite{25}:
\begin{equation}
g_{0}=\sum\limits_{j=1}^N\nabla \lambda_j, \label{3.4}
\end{equation}
which connects the potentials $u$ and $v$ with the eigenfunctions $p$ and $q$,
\begin{equation}
u=-2\langle\Lambda^2 p,p\rangle\langle\Lambda p,p\rangle^{-3},\qquad v=\langle\Lambda p,p\rangle^{-2}.
\label{3.5}
\end{equation}
By direct calculation using Eqs. (\ref{3.1}) and (\ref{3.5}), we have
\begin{equation}\label{3.6}\begin{array}{c}
(2\alpha\partial+\frac12\partial^3)\langle p,p\rangle=0,\\ (2\alpha\partial+\frac12\partial^3)\langle\Lambda^kp,p\rangle+
(\partial u+u\partial)\langle\Lambda^{k+1}p,p\rangle+(\partial v+v\partial)\langle\Lambda^{k+2}p,p\rangle=0,\quad k\in \mathbb{Z}.\\
\end{array}
\end{equation}
From Eqs. (\ref{2.3}) and (\ref{3.6}), we define
\begin{equation}\label{3.7}
b_{-1}=-\langle p,p\rangle,\quad b_{-j}=-\langle\Lambda^{-j+1}p,p\rangle,\quad b_j=-\langle\Lambda^{j-1}p,p\rangle,\qquad j\geq2,
\end{equation}
which, together with Eqs. (\ref{2.4}) and (\ref{3.1}), yield the geometric conditions
\begin{equation}\label{3.8}
\langle p,p\rangle=1,\qquad \langle p,q\rangle=0.
\end{equation}

\par
Simply substituting Eqs. (\ref{3.5}), (\ref{3.7}), and (\ref{3.8}) back into the Lax representations (\ref{2.1}) and (\ref{2.15}), we arrive at two Neumann type systems in the vector form
\begin{equation}\label{3.9}\left\{\begin{array}{lll}
p_x&=&q,\\
q_x&=&-\alpha p+\frac{2\langle\Lambda^2 p,p\rangle}{\langle\Lambda p,p\rangle^3}\Lambda p-\frac{1}{\langle\Lambda p,p\rangle^2}\Lambda^2 p,\\
&& \langle p,p\rangle=1,\quad \langle p,q\rangle=0,
\end{array}\right.
\end{equation}
and
\begin{equation}\label{3.10}\left\{\begin{array}{lll}
p_{t_{-3}}&=&-\langle\Lambda^{-1}p,q\rangle p+\langle\Lambda^{-1}p,p\rangle q+\Lambda^{-1}q,\\
q_{t_{-3}}&=&-\left(\langle\Lambda^{-1}q,q\rangle-\langle\Lambda p,p\rangle^{-1}\right)p-\alpha\Lambda^{-1}p+
2\langle\Lambda^2 p,p\rangle\langle\Lambda^{-1}p,p\rangle\langle\Lambda p,p\rangle^{-3}\Lambda p\\&&
-\langle\Lambda p,p\rangle^{-2}\Lambda p-\langle\Lambda^{-1}p,p\rangle\langle\Lambda p,p\rangle^{-2}\Lambda^2p+\langle\Lambda^{-1}p,q\rangle q,\\&&
\langle p,p\rangle=1,\quad \langle p,q\rangle=0,
\end{array}\right.
\end{equation}
where we have used the identity
\begin{equation}\label{3.17}
\langle q,q\rangle+\langle\Lambda^2p,p\rangle\langle\Lambda p,p\rangle^{-2}=\alpha,
\end{equation}
which follows from the second derivative of the sphere condition $\langle p,p\rangle=1$ in variable $x$. Actually, the left-hand side of Eq. (\ref{3.17}) is a constant of motion [see Eq. (\ref{3.28}) below], which makes sense in the process of nonlinearization of Lax pairs. Because the eigenfunctions $p$ and $q$ belong to $\mathbb{R}^{N}$, if the $N$ eigenvalues are also restricted to be real, it is clear that the parameter $\alpha$ is larger than 0. On one hand, if $\alpha$ lies in the interval $(0,1]$, it is found from Eq. (\ref{3.17}) that
\begin{equation}\label{a1}\begin{array}{l}
\sum\limits_{j<k}(\lambda_j^2+\lambda_k^2)p_j^2p_k^2=\sum\limits_{j=1}^N\lambda_j^2p_j^2\sum\limits_{k=1,k\neq j}^Np_k^2
=\sum\limits_{j=1}^N\lambda_j^2p_j^2(1-p_j^2)\\
\leq\sum\limits_{j=1}^N\lambda_j^2p_j^2(\frac{1}{\alpha}-p_j^2)\leq2\sum\limits_{j<k}\lambda_j\lambda_kp_j^2p_k^2.
\end{array}\end{equation}
This means that the inequality (\ref{a1}) holds true only for $\lambda_j=\lambda_k$, which contradicts the assumption of distinct nonzero eigenvalues. On the other hand, if $\alpha>1$,
\begin{equation}\label{a2}
\sum\limits_{j=1}^N\lambda_j^2p_j^2(1-\alpha p_j^2)\leq\sum\limits_{j=1}^N\lambda_j^2p_j^2(1-p_j^2)=\sum\limits_{j<k}(\lambda_j^2+\lambda_k^2)p_j^2p_k^2\leq
2\alpha\sum\limits_{j<k}\lambda_j\lambda_kp_j^2p_k^2,
\end{equation}
which means that the Neumann type systems (\ref{3.9}) and (\ref{3.10}) may have some solutions in $\mathbb{R}^{2N}$.

Setting $n=2$ in Eq. (\ref{2.11}), it is seen from Eq. (\ref{2.4}) that $V^{(-2)}$ is in fact the spectral matrix $U$. Therefore, the finite-dimensional dynamical systems (\ref{3.9}) and (\ref{3.10}) can be regarded as the backward Neumann type systems because the Lax pair (\ref{2.1}) and (\ref{2.15}) belong to the negative Lax representations (\ref{2.11}).
Note that the backward Neumann type systems (\ref{3.9}) and (\ref{3.10}) admit the common geometric condition (\ref{3.8}) restricting themselves on the tangent bundle of a unit sphere
\begin{equation}\label{3.11}
TS^{N-1}=\{\langle p,q\rangle\in \mathbb{R}^{2N}| F=:\langle p,p\rangle-1=0,\ G=:\langle p,q\rangle=0\},
\end{equation}
we consider the Dirac-Poisson bracket
\begin{equation}\label{3.12}
\{f,g\}_{D}=\{f,g\}+\frac{1}{\{F,G\}}(\{f,F\}\{G,g\}-\{f,G\}\{F,g\})
\end{equation}
and rewrite the backward Neumann type systems (\ref{3.9}) and (\ref{3.10}) as the canonical Hamiltonian equations
\begin{equation}\label{3.13}
p_x=\{p, H_0\}_{D},\qquad q_x=\{q, H_0\}_{D},
\end{equation}
\begin{equation}\label{3.14}
p_{t_{-3}}=\{p, H_{-1}\}_{D},\qquad q_{t_{-3}}=\{q, H_{-1}\}_{D},
\end{equation}
where
\begin{equation}\label{3.15}
H_0=-\frac12(\langle q,q\rangle+\langle\Lambda p,p\rangle^{-2}\langle\Lambda^2 p,p\rangle),
\end{equation}
\begin{equation}\label{3.16}
H_{-1}=-\frac12(\langle\Lambda^2 p,p\rangle\langle\Lambda^{-1} p,p\rangle\langle\Lambda p,p\rangle^{-2}-\langle\Lambda p,p\rangle^{-1}
+\langle\Lambda^{-1} q,q\rangle+\langle\Lambda^{-1} p,p\rangle\langle q,q\rangle).
\end{equation}
Clearly, it follows from Eqs. (\ref{3.13})-(\ref{3.16}) that the backward Neumann type systems (\ref{3.9}) and (\ref{3.10}) are indeed constrained Hamiltonian systems over the tangent bundle of a unit sphere.

\par
The ncHD equation (\ref{2.14}) has thus been reduced to a pair of backward Neumann type systems (\ref{3.9}) and (\ref{3.10}) on $(TS^{N-1},\omega^2)$ since it is the compatibility condition of Lax representations (\ref{2.1}) and (\ref{2.15}). 
One of the important issues in the theory of integrable systems is to solve high-order INLEEs with explicit solutions. Attention is therefore next paid to Neumann type integrable reduction for the ncHD hierarchy (\ref{2.13}) with $n\geq4$. It is known from Eqs. (\ref{2.5}) and (\ref{3.7}) that
\begin{equation}\label{3.18}
g_{-j}=(\langle\Lambda^{-j+3}p,p\rangle,\langle\Lambda^{-j+4}p,p\rangle)^T,\qquad j\geq4.
\end{equation}
Substituting Eqs. (\ref{3.5}), (\ref{3.8}), and (\ref{3.18}) into the Lax representation (\ref{2.11}) gives rise to the sequence of backward Neumann type systems
\begin{equation}\label{3.19}\left\{\begin{array}{l}
p_{t_{-k-2}}=\frac12\sum\limits_{j=0}^k(2\langle\Lambda^{-j}p,p\rangle\Lambda^{j-k}q-\langle\Lambda^{-j}p,q\rangle\Lambda^{j-k}p
-\langle\Lambda^{j-k}p,q\rangle\Lambda^{-j}p),\\
q_{t_{-k-2}}=-\langle\Lambda^{-k}q,q\rangle p+\langle\Lambda p,p\rangle^{-1}\Lambda^{1-k}p-
\langle\Lambda^{1-k}p,p\rangle\langle\Lambda p,p\rangle^{-2}\Lambda p\\
-\langle\Lambda p,p\rangle^{-2}(\langle\Lambda^{-k}p,p\rangle\Lambda^2p+\langle\Lambda^2 p,p\rangle\Lambda^{-k}p)
+2\langle\Lambda^2 p,p\rangle\langle\Lambda^{-k}p,p\rangle\langle\Lambda p,p\rangle^{-3}\Lambda p\\
-\frac12\sum\limits_{j=0}^k(2\langle\Lambda^{j-k}q,q\rangle\Lambda^{-j}p-\langle\Lambda^{-j}p,q\rangle\Lambda^{j-k}q
-\langle\Lambda^{j-k}p,q\rangle\Lambda^{-j}q),\\
\langle p,p\rangle=1,\quad \langle p,q\rangle=0,\qquad k\geq2.\\
\end{array}\right.
\end{equation}
Moreover, using the Dirac-Poisson bracket and the identity (\ref{3.17}), the backward Neumann type systems (\ref{3.19}) can also be expressed as the canonical Hamiltonian equation
\begin{equation}\label{3.20}
p_{t_{-k-2}}=\{p, H_{-k}\}_{D},\quad q_{t_{-k-2}}=\{q,
H_{-k}\}_{D},\qquad k\geq2,
\end{equation}
where
\begin{equation}\label{3.21}\begin{array}{lll}
H_{-k}&=&\frac12(\langle\Lambda^{1-k}p,p\rangle\langle\Lambda p,p\rangle^{-1}-\langle\Lambda^{-k}p,p\rangle
(\langle\Lambda^2 p,p\rangle\langle\Lambda p,p\rangle^{-2}+\langle q,q\rangle)\\&&
-\langle\Lambda^{-k}q,q\rangle-\sum\limits_{j=1}^{k-1}(\langle\Lambda^{-j}p,p\rangle\langle\Lambda^{j-k}q,q\rangle-
\langle\Lambda^{-j}p,q\rangle\langle\Lambda^{j-k}p,q\rangle)).\\
\end{array}
\end{equation}

\par
Note that each INLEE in the ncHD hierarchy has been reduced to two backward Neumann type systems. The problem of obtaining solutions to the ncHD hierarchy (\ref{2.13}) is converted to dealing with a sequence of backward Neumann type systems (\ref{3.9}), (\ref{3.10}), and (\ref{3.19}). To solve the associated backward Neumann type systems, it is necessary for us to evaluate their integrals of motion. Let us introduce the bilinear generating function
\begin{equation}\label{3.22}
G_\lambda=\left(\begin{array}{c}
0\\
\langle\Lambda p,p\rangle
\end{array}\right)
+\sum\limits_{j=1}^N\frac{\nabla\lambda_j}{\lambda-\lambda_j}=
\left(\begin{array}{c}
Q_\lambda(\Lambda p,p)\\
\langle\Lambda p,p\rangle+Q_\lambda(\Lambda^2 p,p)\\
\end{array}\right),
\end{equation}
which also satisfies the Lenard eigenvalue equation corresponding to $\lambda$,
\begin{equation}\label{3.23}(K-\lambda J)G_\lambda=0,\end{equation}
where
$$Q_\lambda(\xi,\eta)=\sum_{j=1}^{N}\frac{\xi_j\eta_j}
{\lambda-\lambda_j}=-\sum_{m=0}^{\infty}\langle\Lambda^{-m-1}\xi,\eta\rangle\lambda^{m},\quad |\lambda|<\min\{|\lambda_1|,|\lambda_2|,\cdots,|\lambda_N|\}.$$
Simply substituting the expression for $G_\lambda$ into the spectral matrix $V^{(-n)}$, we attain the Lax matrix
\begin{equation}\label{3.24}
V_\lambda=\left(\begin{array}{cc}
-Q_\lambda(\Lambda p,q)&Q_\lambda(\Lambda p,p)\\
\frac{\langle\Lambda^2p,p\rangle}{\langle\Lambda p,p\rangle^2}-\frac{\lambda}{\langle\Lambda p,p\rangle}
-Q_\lambda(\Lambda q,q)&Q_\lambda(\Lambda p,q)\\
\end{array}\right),
\end{equation}
which satisfies the Lax equation
\begin{equation}\label{3.25}
(V_\lambda)_x-[U,V_\lambda]=0,
\end{equation}
using the identities (\ref{2.12}) and (\ref{3.23}). It clearly follows from \cite{35} that the determinant of $V_\lambda$ in Eq. (\ref{3.26}) is the generating function of conserved integrals for the backward Neumann type system (\ref{3.9}) with $|\lambda|<\min\{|\lambda_1|,|\lambda_2|,\cdots,|\lambda_N|\}$,
\begin{equation}\label{3.26}\begin{array}{lll}
\det V_\lambda&=:&F^{-}_\lambda=Q_\lambda(\Lambda p,p)\left(\frac{\lambda}{\langle\Lambda p,p\rangle}-
\frac{\langle\Lambda^2p,p\rangle}{\langle\Lambda p,p\rangle^2}\right)+Q_\lambda(\Lambda p,p)Q_\lambda(\Lambda q,q)-Q^2_\lambda(\Lambda p,q)\\
&=&1+\sum\limits_{j=1}^N\frac{G_j}{\lambda-\lambda_j}=1-\sum\limits_{k=0}^\infty\left(\sum\limits_{j=1}^N\lambda_j^{-k-1}G_j\right)\lambda^k
=F_0+\sum\limits_{k=1}^\infty F_{-k}\lambda^k,\\
\end{array}
\end{equation}
where
\begin{equation}\label{3.27}
G_j=\frac{\lambda_j^2p_j^2}{\langle\Lambda p,p\rangle}-\frac{\langle\Lambda^2 p,p\rangle}{\langle\Lambda p,p\rangle^2}\lambda_jp_j^2
+\sum_{k=1,k\neq j}^N\frac{\lambda_j\lambda_k
(p_jq_k-p_kq_j)^2}{\lambda_j-\lambda_k},\quad 1\leq j\leq N
\end{equation}
and
\begin{eqnarray}
F_0&=&\langle\Lambda^2 p,p\rangle\langle\Lambda p,p\rangle^{-2}+\langle q,q\rangle,\label{3.28}\\
F_{-1}&=&-\langle\Lambda p,p\rangle^{-1}+\langle\Lambda^{-1} p,p\rangle\langle\Lambda^2 p,p\rangle\langle\Lambda p,p\rangle^{-2}+
\langle\Lambda^{-1} q,q\rangle+\langle\Lambda^{-1} p,p\rangle\langle q,q\rangle,\label{3.29}\\
F_{-k}&=&-\frac{\langle\Lambda^{1-k} p,p\rangle}{\langle\Lambda p,p\rangle}+\frac{\langle\Lambda^2 p,p\rangle}{\langle\Lambda p,p\rangle^{2}}
\langle\Lambda^{-k} p,p\rangle+\sum\limits_{j=0}^k
\left|\begin{array}{cc}
\langle\Lambda^{-j}p,p\rangle&\langle\Lambda^{j-k}p,q\rangle\\
\langle\Lambda^{-j}p,q\rangle&\langle\Lambda^{j-k}q,q\rangle \\
\end{array}\right|,\ k\geq2.\label{3.30}
\end{eqnarray}
\par
Actually, $\{F_{-k}\}$ are also the integrals of motion for the backward Neumann type systems (\ref{3.10}) and (\ref{3.19}) [see Eq. (\ref{4.6})]. By considering the integrals of motion given by Eqs. (\ref{3.28})-(\ref{3.30}), it will be seen in Sec. \ref{sec:4} that the backward Neumann type flows commute with each other over $(TS^{N-1},\omega^2)$ and that the Hamiltonian vector fields (\ref{3.13}), (\ref{3.14}), and (\ref{3.20}) are exactly mapped into the ncHD vector fields (\ref{2.13}) via the symmetric constraint (\ref{3.5}).

\setcounter{equation}{0}
\section{From the Neumann Type Systems to the ncHD Hierarchy}
\label{sec:4}
Forward FDISs have been adapted to solve positive-order INLEEs \cite{27,28,29,30,32,33-1}. Therein, the key step is that the relation between the positive-order INLEEs and the forward FDISs is established, where the symmetric constraint specifies a finite-dimensional invariant subspace for the positive-order flows. In this section, we establish the relation between the negative-order INLEEs and the backward Neumann type systems.

\par
To progress further, we denote the flow variables of $F^-_\lambda$, $F_{-k} \ (k\geq0)$, and $H_{-k}\ (k\geq0)$ by $\tau^-_\lambda$, $\tau_{-k} \ (k\geq0)$, and $t_{-k-2}\ (k\geq0)$, respectively, and we regard $F^-_\lambda$ as the Hamiltonian on $(TS^{N-1},\omega^2)$. A direct calculation leads to the following canonical Hamiltonian equation in the Dirac-Poisson bracket:
\begin{equation}
\frac{d}{d\tau^{-}_\lambda}\left(\begin{array}{c}
p_k\\
q_k\\
\end{array}\right)
=\left(\begin{array}{c}
\{p_k,F^{-}_\lambda\}_{D}\\
\{q_k,F^{-}_\lambda\}_{D}\\
\end{array}\right)=W(\lambda,\lambda_k)\left(\begin{array}{c}
p_k\\
q_k\\
\end{array}\right) ,\label{4.1}
\end{equation}
where
\begin{equation}
W(\lambda,\mu)=-\frac{2\mu}{\lambda-\mu}V_\lambda+\frac{2\mu Q_\lambda(\Lambda p,p)}{\langle\Lambda p,p\rangle^2}\left(
\frac{2\langle\Lambda^2 p,p\rangle}{\langle\Lambda p,p\rangle}-\mu-\lambda\right)
\sigma_3.\label{4.2}\end{equation}
\begin{proposition}
  \label{prop:4.1}
On $(TS^{N-1},\omega^2)$, the Lax matrix $V_\mu$ satisfies the Lax equation
\begin{equation}
\frac{dV_\mu}{d\tau^-_\lambda}=[W(\lambda,\mu)
, V_\mu],\qquad \lambda\neq\mu, \quad \lambda,\ \mu\in\mathbb{C}. \label{4.3}
\end{equation}
 \end{proposition}
{\bf Proof:} For the convenience of description, we rewrite the Lax matrix $V_\mu$ as
$$
V_\mu=\left(\frac{\langle\Lambda^2p,p\rangle}{\langle\Lambda p,p\rangle^2}-\frac{\mu}{\langle\Lambda p,p\rangle}
\right)\sigma_3+\sum\limits_{k=1}^N\frac{\epsilon_k}{\mu-\lambda_k},
$$
where $\epsilon_k=-\lambda_k(p_kq_k\sigma_1-p_k^2\sigma_2+q_k^2\sigma_3).$ Starting from Eq. (\ref{4.1}), after a direct calculation we arrive at
$$\frac{d\epsilon_k}{d\tau^-_\lambda}=[W(\lambda,\lambda_k),\epsilon_k],\quad
\frac{d}{d\tau^-_\lambda}\left(\frac{\langle\Lambda^2p,p\rangle}{\langle\Lambda p,p\rangle^2}-\frac{\mu}{\langle\Lambda p,p\rangle}\right)
=\kappa,
$$where
\begin{eqnarray*}\kappa&=&4Q_\lambda(\Lambda p,q)\left(\frac{\langle\Lambda^2p,p\rangle}{\langle\Lambda p,p\rangle^2}-\frac{\lambda+\mu}{\langle\Lambda p,p\rangle}\right)+\frac{4Q_\lambda(p,p)}{\langle\Lambda p,p\rangle^2}(\lambda\langle\Lambda p,q\rangle+\langle\Lambda^2p,q\rangle)\\&&
+4\langle\Lambda p,q\rangle Q_\lambda(\Lambda p,p)\left(\frac{\mu}{\langle\Lambda p,p\rangle^2}-
\frac{2\langle\Lambda^2p,p\rangle}{\langle\Lambda p,p\rangle^3}
\right).
\end{eqnarray*}
Then, we arrive at
\begin{eqnarray*}
\frac{dV_\mu}{d\tau^-_\lambda}&=&\kappa\sigma_3+\sum_{k=1}^{N}\frac{1}{\mu-\lambda_k}[W(\lambda,\lambda_k),\epsilon_k]\\
&=&\kappa\sigma_3+\frac{Q_\lambda(\Lambda p,p)}{\langle\Lambda p,p\rangle^2}\sum_{k=1}^{N}\frac{2\lambda_k}{\mu-\lambda_k}\left[\left(
\frac{2\langle\Lambda^2 p,p\rangle}{\langle\Lambda p,p\rangle}-\lambda_k-\lambda\right)\sigma_3,\epsilon_k\right]\\&&
-\frac{2}{\lambda-\mu}\sum_{k=1}^{N}\left(\frac{1}{\mu-\lambda_k}
-\frac{1}{\lambda-\lambda_k}\right)[V_\lambda,\lambda_k\epsilon_k]\\
&=&\kappa\sigma_3-2Q_\lambda(\Lambda p,p)\left(
\frac{2\langle\Lambda^2 p,p\rangle}{\langle\Lambda p,p\rangle^3}-\frac{\lambda}{\langle\Lambda p,p\rangle^2}\right)
(Q_\mu(\Lambda^2p,p)\sigma_1+2Q_\mu(\Lambda^2p,q)\sigma_3)\\&&
+\frac{2Q_\lambda(\Lambda p,p)}{\langle\Lambda p,p\rangle^2}(Q_\mu(\Lambda^3p,p)\sigma_1+2Q_\mu(\Lambda^3p,q)\sigma_3)
-\frac{2}{\lambda-\mu}[V_\lambda,\mu V_\mu]\\&&
+2\left(\frac{\lambda+\mu}{\langle\Lambda p,p\rangle}-\frac{\langle\Lambda^2 p,p\rangle}{\langle\Lambda p,p\rangle^2}
\right)(Q_\lambda(\Lambda p,p)\sigma_1+2Q_\lambda(\Lambda p,q)\sigma_3)\\
&=&\frac{2\mu Q_\lambda(\Lambda p,p)}{\langle\Lambda p,p\rangle^2}\left(
\mu+\lambda-\frac{2\langle\Lambda^2 p,p\rangle}{\langle\Lambda p,p\rangle}\right)
(Q_\mu(\Lambda p,p)\sigma_1+2Q_\mu(\Lambda p,q)\sigma_3)-\frac{2\mu}{\lambda-\mu}[V_\lambda,V_\mu]\\
&=&[W(\lambda,\mu), V_\mu],
\end{eqnarray*}
by using the identity
$$Q_\lambda(\Lambda^l\xi,\eta)=\lambda
Q_\lambda(\Lambda^{l-1}\xi,\eta)+\langle\Lambda^{l-1}\xi,\eta\rangle,\qquad l\in\mathbb{Z},$$
which completes the proof.
\begin{corollary}
  \label{coro:4.1}
\begin{equation}\label{4.4}
\{F^-_\mu,F^-_\lambda\}_{D}=0,\qquad \forall \lambda,\ \mu\in\mathbb{C},
\end{equation}
\begin{equation}\label{4.5}
\{F_{-j},F_{-k}\}_{D}=0,\qquad j,\ k=0,1,2,\cdots .
\end{equation}
\end{corollary}

\par
Recalling Eqs. (\ref{3.15}), (\ref{3.16}), (\ref{3.21}), and (\ref{3.28})-(\ref{3.30}), it is found that $H_{-k}=-\frac12F_{-k}$, $k\geq0$.
From the Dirac-Poisson bracket, we have
\begin{equation}\label{4.6}
\frac{dF_{-j}}{dt_{-k-2}}=\{F_{-j},H_{-k}\}_D=-\frac12\{F_{-j},F_{-k}\}_D=0,\qquad j,\ k=0,1,2,\cdots,
\end{equation}
which indicates that $\{F_{-k}\}$ $(k\geq0)$ are also the conserved integrals for the backward Neumann type systems (\ref{3.10}) and (\ref{3.19}). Note that the conserved integral $G_{j}$ corresponds to the nonzero eigenvalue $\lambda_j$. The choice of distinct eigenvalues ensures that $\{F_0,F_{-1},\cdots,F_{-N+2}\}$ are functionally independent on $(TS^{N-1},\omega^2)$. Therefore, we can show the Liouville integrability for the backward Neumann type systems.
\begin{proposition}
  \label{prop:4.2}
The backward Neumann type systems (\ref{3.9}), (\ref{3.10}), and (\ref{3.19}) are completely integrable in the Liouville sense.
\end{proposition}

\par
It follows from Proposition \ref{prop:4.2} that the backward Neumann type flows mutually commute over $(TS^{N-1},\omega^2)$. Thus, there exists a smooth function in the variables of $t_{-j-2}$ and $t_{-k-2}$ $(j,\ k\geq0)$ giving an involutive solution to the backward Neumann type systems $(H_{-j},\omega^2,TS^{N-1})$ and $(H_{-k},\omega^2,TS^{N-1})$ \cite{36}. Owing to the commutability of backward Neumann type flows, we arrive at the following propositions.
\begin{proposition}\label{prop:4.3}
Let $(p(x,t_{-3}),q(x,t_{-3}))^T$ be the involutive solution of backward Neumann type systems (\ref{3.9}) and (\ref{3.10}). Then,
\begin{equation}\begin{array}{l}
u=-\frac{2\langle\Lambda^2 p(x,t_{-3})),p(x,t_{-3}))\rangle}{\langle\Lambda p(x,t_{-3})),p(x,t_{-3}))\rangle^{3}},\quad
 v=\frac{1}{\langle\Lambda p(x,t_{-3})),p(x,t_{-3}))\rangle^{2}}, \\ w=-\langle\Lambda^{-1}p(x,t_{-3}),p(x,t_{-3})\rangle
 \end{array}
\label{4.7}
\end{equation}
solves the ncHD equation (\ref{2.14}).
 \end{proposition}
{\bf Proof:} Employing the backward Neumann type system (\ref{3.9}), we obtain
\begin{equation}\label{4.8}
u_x=\frac{12\langle\Lambda^2p,p\rangle\langle\Lambda p,q\rangle}{\langle\Lambda p,p\rangle^4}-\frac{4\langle\Lambda^2 p,q\rangle}{\langle\Lambda p,p\rangle^3},\quad v_x=-\frac{4\langle\Lambda p,q\rangle}{\langle\Lambda p,p\rangle^3},
\end{equation}
and
\begin{equation}\label{4.8b}
w_x=-2\langle\Lambda^{-1} p,q\rangle,\quad
w_{xx}=2\left(u+\alpha\langle\Lambda^{-1}p,p\rangle+\frac{1}{\langle\Lambda p,p\rangle}-\langle\Lambda^{-1}q,q\rangle\right).
\end{equation}
On the other hand, from the backward Neumann type system (\ref{3.10}), we have
\begin{equation}\label{4.9}
\langle\Lambda p,p\rangle_{t_{-3}}=2(\langle\Lambda^{-1} p,p\rangle\langle\Lambda p,q\rangle-\langle\Lambda^{-1} p,q\rangle\langle\Lambda p,p\rangle),
\end{equation}
\begin{equation}\label{4.10}
\langle\Lambda^2 p,p\rangle_{t_{-3}}=2(\langle\Lambda p,q\rangle+\langle\Lambda^{-1} p,p\rangle\langle\Lambda^2 p,q\rangle-
\langle\Lambda^{-1} p,q\rangle\langle\Lambda^2 p,p\rangle).
\end{equation}
It follows from Eqs. (\ref{4.8})-(\ref{4.10}) that Eq. (\ref{4.7}) satisfies the first two equations in Eq. (\ref{2.14}). Substituting Eq. (\ref{4.8b}) back into the third equation of Eq. (\ref{2.14}), we have
\begin{eqnarray}
u&=&u+\frac{1}{\langle\Lambda p,p\rangle}-\alpha\langle\Lambda^{-1} p,p\rangle-\langle\Lambda^{-1}q,q\rangle\nonumber\\
&=&u+\frac{1}{\langle\Lambda p,p\rangle}-\langle\Lambda^{-1} p,p\rangle\left(
\frac{\langle\Lambda^2 p,p\rangle}{\langle\Lambda p,p\rangle^2}+\langle q,q\rangle
\right)-\langle\Lambda^{-1}q,q\rangle,\label{cc}
\end{eqnarray}
by using Eq. (\ref{3.17}). It follows from Eq. (\ref{3.29}) that Eq. (\ref{4.7}) also satisfies the third equation in Eq. (\ref{2.14}), from which the redundant term in Eq. (\ref{cc}) is the exact constant of motion (\ref{3.29}) and hence can be set to zero.

\begin{proposition}\label{prop:4.4}
Let $(p(x,t_{-n-2}), q(x,t_{-n-2}))^T \ (n\geq2)$ be an involutive solution of
backward Neumann type systems (\ref{3.9}) and (\ref{3.19}). Then
\begin{equation}
u=-\frac{2\langle\Lambda^2 p(x,t_{-n-2})),p(x,t_{-n-2}))\rangle}{\langle\Lambda p(x,t_{-n-2})),p(x,t_{-n-2}))\rangle^{3}},\quad
v=\frac{1}{\langle\Lambda p(x,t_{-n-2})),p(x,t_{-n-2}))\rangle^{2}}
\label{4.12}
\end{equation}
are the finite parametric solutions to the $n$th ncHD equation (\ref{2.13}).
\end{proposition}
{\bf Proof:} Referring to the backward Neumann type systems (\ref{3.19}), a direct computation yields
\begin{equation}\label{4.13}
\langle\Lambda p,p\rangle_{t_{-n-2}}=
2(\langle\Lambda^{-n} p,p\rangle\langle\Lambda p,q\rangle-\langle\Lambda^{-n} p,q\rangle\langle\Lambda p,p\rangle),
\end{equation}
\begin{equation}\label{4.14}\begin{array}{lll}
\langle\Lambda^2 p,p\rangle_{t_{-n-2}}&=&2(\langle\Lambda^{-n+1} p,p\rangle\langle\Lambda p,q\rangle-
\langle\Lambda^{-n+1} p,q\rangle\langle\Lambda p,p\rangle\\&&+
\langle\Lambda^{-n} p,p\rangle\langle\Lambda^2 p,q\rangle-
\langle\Lambda^{-n} p,q\rangle\langle\Lambda^2 p,p\rangle).
\end{array}
\end{equation}
From the backward Neumann type systems (\ref{3.9}), a similar calculation gives
\begin{equation}\label{4.15}
(\partial v+v\partial)\langle\Lambda^{-n}p,p\rangle=4\left(
\frac{\langle\Lambda^{-n}p,q\rangle}{\langle\Lambda p,p\rangle^2}-\frac{\langle\Lambda p,q\rangle\langle\Lambda^{-n}p,p\rangle}
{\langle\Lambda p,p\rangle^3}
\right),
\end{equation}
and
\begin{equation}\label{4.16}\begin{array}{l}
-(2\alpha\partial+\frac12\partial^3)\langle\Lambda^{-n-1}p,p\rangle=4\langle\Lambda^{-n}p,p\rangle\left(
\frac{3\langle\Lambda^2 p,p\rangle\langle\Lambda p,q\rangle}{\langle\Lambda p,p\rangle^4}-\frac{\langle\Lambda^2 p,q\rangle}{\langle\Lambda p,p\rangle^3}\right)\\
-\frac{4}{\langle\Lambda p,p\rangle^3}\left(2\langle\Lambda^2 p,p\rangle\langle\Lambda^{-n} p,q\rangle+
\langle\Lambda p,q\rangle\langle\Lambda^{-n+1} p,p\rangle\right)+
\frac{4\langle\Lambda^{-n+1} p,q\rangle}{\langle\Lambda p,p\rangle^2}.
\end{array}\end{equation}
By using Eqs. (\ref{2.6}) and (\ref{3.18}), the combination of Eqs. (\ref{4.13})-(\ref{4.16}) indicates that Eq. (\ref{4.12}) solves the $n$th ncHD equation (\ref{2.13}).

\par
To sum up, on the basis of Propositions \ref{prop:4.3} and \ref{prop:4.4}, the ncHD hierarchy with $\alpha>1$ has been subjected to finite-dimensional integrable reduction on the tangent bundle of a unit sphere. It is found that a nonlocal INLEE can be reduced to a pair of local FDISs by separating the temporal and spatial variables. Moreover, the relation between the negative-order integrable systems and the backward Neumann type systems is specified, where the involutive solutions of backward Neumann type systems generate finite parametric solutions of negative-order INLEEs. As a consequence, the construction of explicit solutions to negative-order INLEEs is simplified to solving some FDISs of solvable ordinary differential equations under a symmetric constraint. Referring to the application of forward FDISs \cite{27,28,29,30,32,33-1}, one may take the backward FDISs as a basis for obtaining explicit solutions of nonlocal INLEEs. Based on the results given in this paper, we hope to solve some negative-order INLEEs of physical interest by means of the backward Neumann type systems.

\section*{Acknowledgment}
This work was supported by the National Natural Science Foundation
of China (No.11471072).

\end{document}